\documentclass[useAMS,usenatbib]{mn2e}
\usepackage{graphicx}

\def\ergscm2 {erg\,s$^{-1}$cm$^{-2}$}

\def\ee {1E\,1048.1--5937\,}
\def\kes {1E\,1841--045\,}
\def\axj {AX\,J1845--0258\,}
\def\rxj {1RXS\,J170849.0--400910\,}

\def\xte {XTE\,J1810--197\,}

\def\newpsr{PSR\,J1712--3943}
\def\ltsim{\raisebox{-.5ex}{$\;\stackrel{<}{\sim}\;$}}

\title[Search for radio pulsations in 4 AXPs and discovery of two
  new pulsars]{Search for radio pulsations in four Anomalous X-ray
  Pulsars and discovery of two new pulsars}

\author[M. Burgay, N. Rea, G.L. Israel, A. Possenti, L. Burderi, T. Di Salvo,
  N. D'Amico, L. Stella]{M. Burgay$^{1}$\thanks{E-mail:
    burgay@ca.astro.it}, N. Rea$^{2}$, G.L. Israel$^{3}$,
  A. Possenti$^{1}$, L. Burderi$^{4}$, \newauthor  T. Di Salvo$^{5}$, N. D'Amico$^{4}$, L. Stella$^{3}$ \\
$^{1}$INAF - Osservatorio Astronomico di Cagliari, Loc. Poggio dei Pini, Strada 54, 09012 Capoterra (CA),Italy \\
$^{2}$SRON - Netherlands Institute for Space Research, Sorbonnelaan, 2, 3584 CA, Utrecht, The Netherlands \\
$^{3}$INAF - Osservatorio Astronomico di Roma, via di Frascati 33, Monte Porzio Catone, Roma, Italy \\
$^{4}$Dipartimento di Fisica dell'Universit\`a di Cagliari, S.P. Monserrato-Sestu Km 0,700, 09042 Monserrato (CA), Italy \\
$^{5}$Dipartimento di Scienze Fisiche e Astronomiche dell'Universit\`a di Palermo, via Archirafi 36, 90134 Palermo, Italy\\
}

\begin{document}

\date{Draft: 2006 June 1st}

\pagerange{\pageref{firstpage}--\pageref{lastpage}} \pubyear{2006}

\maketitle

\label{firstpage}

\begin{abstract}
We report on observations of four southern Anomalous X-ray Pulsars,
(\rxj, \ee, \kes and \axj), obtained at 1.4 GHz using the Parkes radio
telescope. Radio pulsations from these sources have been searched (i)
by directly folding the time series at a number of trial periods
centered on the value of the spin rate obtained from the X-ray
observations; (ii) by performing a blind search; (iii) using a code
sensitive to single dedispersed pulses, in the aim to detect signals
similar to those of the recently discovered Rotating RAdio
Transients. No evidence for radio pulsations with an upper limit of
$\sim 0.1$\,mJy for any of the four targets has been found. The blind
search led to the serendipitous discovery of two new pulsars, rotating
with a spin period of about 0.7\,s and of 92\,ms respectively, and to
the further detection of 18 known pulsars, two of which were also
detected in the single-pulse search.
\end{abstract}


\section{Introduction}

In the last two decades, a significant amount of observational and
theoretical effort has been dedicated towards the understanding of an
``unusual'' class of pulsars, namely the Anomalous X-Ray Pulsars (AXPs)
and the Soft Gamma-Ray Repeaters (SGRs). These relatively bright X-ray
sources ($10^{34}-10^{36}$\,erg\,s$^{-1}$), discovered either as
peculiar short X-ray bursts or steady X-ray emitters, were soon
recognised as a distinct class of neutron stars with respect to the
well known radio pulsar or X-ray binary populations (see \citealt{wt04}
for a review).

The arguments that lead to this conclusion are the following: first,
AXPs and SGRs are X-ray pulsators with spin periods in the range
5-12\,s. Furthermore, their rotational energy loss $\dot{E}$ inferred
from their spin-down rate ($\dot{E}=I\Omega\dot{\Omega}$, where I,
$\Omega$ and $\dot{\Omega}$ are respectively the neutron star moment
of inertia, angular velocity and angular velocity time derivative) is
insufficient to power the observed X-ray luminosity (note that for the
large majority of isolated pulsars, detected both in radio and X-ray
band, the X-ray luminosity is typically 0.1\% of the $\dot{E}$;
\citealt{bt97,pccm02}). Second, there is no evidence for a companion
star (through the search for Doppler orbital modulation of the X-ray
spin period, and through observations at other wavelengths;
\citealt{mis98,wdf+99,hvk00}), which might power their emission through
the accretion mechanism, at least with a mass greater than
0.1\,M$\odot$. An alternative source of energy is therefore required
to power the X-ray emission of AXPs and SGRs.

Assuming these sources being isolated X-ray pulsars with a dipole
magnetic field configuration and purely magnetic dipole losses,
their inferred magnetic fields are extremely high, B$\sim
3.2\times10^{19}\sqrt{P\dot{P}}\sim 10^{14}-10^{15}$~G: these
magnetic fields have been identified as the possible additional
energy reservoir. In fact, if AXPs and SGRs are indeed
"magnetars", i. e. isolated highly magnetic neutron stars, many of
their bizarre X-ray properties might be explained: the X-ray
luminosity is naturally provided by the decay of these
ultra-strong magnetic fields, while their peculiar short bursts
may result from magnetic induced cracks on the neutron star
surface (\citealt{dt92a,td93a}; 1995). Moreover, the long lasting
flux enhancements observed in some magnetars \citep{roz+05,
mte+05} may be associated to single fractures on a larger spatial
scale, or to the episodic onset of a wind energised by frequent,
small scale fractures and/or quasi-steady seismic vibrations
\citep{tlk02}. In this scenario, the magnetars' magnetic fields
are all above the so-called ``quantum critical field'' $B_c =
4.4\times 10^{13}$~G, at which the radio pulsations are expected
to be suppressed by processes such as photon splitting which may
inhibit pair-production cascades \citep{bh98} responsible for
creating the radio emission of the normal radio pulsars.  However,
the discovery of high magnetic field radio pulsars
\citep{ckl+00,msk+03} and transient radio pulsed emission from one
AXP \citep{crh+06} put many uncertainties with this respect making
again theoretically plausible a link between AXPs and radio
pulsars.

In the past years, radio observations of AXPs and SGRs, performed
using different antennas and arrays, have given always negative
results (e.g. \citealt{cpkm02}) both to timing and interferometry
searches, but revealing in a few cases some reliable associations with
SNRs \citep{gsgv01} and in one case an interesting stellar wind bubble
blown by the massive progenitor \citep{gmo+05}. However, in some
peculiar circumstances radio counterparts were detected: a) the SGRs'
Giant Flares, rare, catastrophic and highly energetic events, have
been discovered powering a radio outburst, which slowly faded with a
timescale of weeks \citep{fkb99,ccr+05,gkg+05}, b) during the outburst
of the only confirmed transient AXP, XTE J1810-197, the Very Large
Array (VLA) observations (with angular resolution of 6") revealed a
point source of flux density $4.5\pm0.5$\,mJy at 1.4 GHz
\citep{hgb+05}, which has recently been discovered as pulsated at the
neutron star spin period \citep{crh+06}.

With this picture in mind we have undertaken a systematic deep search
for radio pulsations in three confirmed (\rxj, \ee, \kes) and one
candidate AXP (\axj), visible from the southern hemisphere using the
Parkes radio telescope. Observations and data analysis are described
in \S\ref{sec:obs} while in \S\ref{sec:results} we report on the
obtained results, discussed in the context of the magnetar scenario in
\S\ref{sec:conclu}.

\begin{figure}
\begin{center}
\includegraphics[width=8truecm, angle=270]{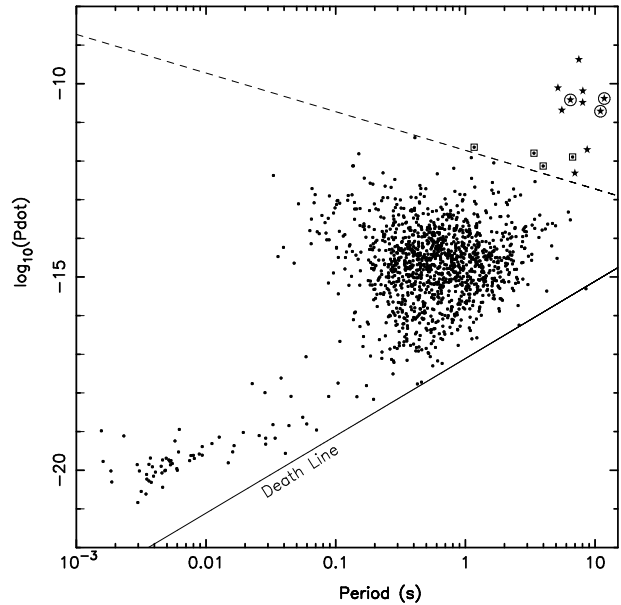}
\caption{$P-\dot{P}$ diagram for radio pulsars (dots) and magnetars
(stars). The dashed line represents the limit for radio emission
quenching. The four sources surrounded by a square emit in radio
besides being above this line. The three confirmed AXPs of our sample
for which a determination of $\dot{P}$ is available are surrounded by
a circle.}
\label{fig1}
\end{center}
\end{figure}

\section{Observations and Data analysis}
\label{sec:obs}

In Table \ref{tab1} the sample of observed AXPs is presented. For each
object the name, period and period derivative of X-ray pulsations, the derived
surface magnetic field, the estimated distance D and the association with a SNR
or with a nebula are listed.

\begin{table*}
\begin{center}
\begin{tabular}{lrccccc}
\hline
Name & $P$ & $\dot{P}$ & B  & D  & DM$_{nom}$ & association \\
     & (s) & (10$^{-11}$\,s\,s$^{-1}$) & (10$^{14}$\,G) & (kpc) & (pc
cm$^{-3}$) & \\
\hline
\ee  &  6.456109(5) & 3.3 & 4.7 & 5 & 279.7 & HI bubble \\
\rxj & 11.00170(4)  & 1.9 & 4.6 & 8 & 742.5 & -- \\
\kes & 11.77505(5)  & 4.1 & 7.0 & 7 & 529.7 & G27.4$+$0.0 \\
\axj &  6.9712(1)  & --  & --   & 8 & 646.1 & G29.6$+$0.1 \\
\hline
\end{tabular}
\caption{For each of the four AXPs of our sample columns 2 and 3
report the most recent period and period derivative from the X-ray
timing (1E1048: \citealt{tmt+05}, 1RXS J1708: \citealt{roz+05}, 1E
1841: \citealt{ggk+02}, AXJ 1845: \citealt{gv98}), column 4 is the
surface magnetic field derived from the dipole formula $B = 3.2 \times
10^{19}\sqrt{P\dot{P}}$, columns 5 and 6 report the distance estimated
and the relative dispersion measure derived adopting the \citet{cl02}
model for the distribution of free electrons in the ISM (the values
obtained using \citet{tc93} model are similar) and last column shows
the association with supernova remnant or with hydrogen bubble
\citep{gsgv01,gmo+05}. Numbers in parentheses are the errors on the
last quoted digit.}
\label{tab1}
\end{center}
\end{table*}

The radio observations have been performed between October 12 and
15 1999 using the 13 beams of the multibeam receiver
\citep{swb+96} of the Parkes radio telescope (NSW Australia) at a
frequency of 1374 MHz, with the central beam pointed on the target
AXP. In order to mitigate the effects of the dispersion of the
signals in the interstellar medium (ISM), the total 288 MHz
bandwidth is splitted into 96 frequency channels each 3-MHz wide.
The outputs from each channel are summed in polarisation pairs,
high-pass filtered and 1-bit sampled every 1.0 or 1.2 ms.  Each
source has been pointed for 2.8 hours and for all but
1E1841$-$0450 the observation has been performed twice.

The data for the central beams have been first analysed using the
{\ttfamily{pdm}} programme: the code takes as input a period P and
a dispersion measure DM and folds the time series according to a
number of trial values around the input ones, searching for the
combination of P and DM for which the signal-to-noise ratio S/N is
maximised. The period range searched for each source has been
obtained from the X-ray ephemeris and their errors. Since at the
time of the data analysis most of the X-ray timing solutions had
big uncertainties, the range explored (see Table \ref{tab2}) is
much wider than would have been necessary with the present
knowledge. A number of trial DM values, ranging from 0 up to the
value DM$_{max}$ giving a maximum pulse broadening in a 3 MHz
frequency channel of 10\% of the pulse period (see Table
\ref{tab2}) has been explored. DM$_{max}$ is well beyond what
expected for a source in the Milky Way on the basis of the
available models for the distribution of free electrons in the
Galaxy \citep{tc93,cl02}, but may take into account the
possibility of the presence of dense local matter.

Making use of the relation derived by \citet{bcc+04} between the
broadening of the pulse due to the scattering caused by the ISM,
$\delta t_{scatt}$ and the DM, we note that already a DM $\sim 1200$
pc cm$^{-3}$ results in a broadening of $\sim 10$ seconds at 1.4
GHz. However the relation by \citet{bcc+04} has a very high scattering
in principle allowing for reasonable values of $\delta t_{scatt}$ also
for much higher DMs. Hence, for the relatively fast search performed
with {\ttfamily{pdm}} we have explored all the DMs up to DM$_{max}$
indicated in Table \ref{tab2}. For the subsequent steps of the
analysis (see later), we have on the contrary decided to adopt less
conservative values for the maximum DM explored.

\begin{center}
\begin{figure}
\includegraphics[angle=270, width=8truecm]{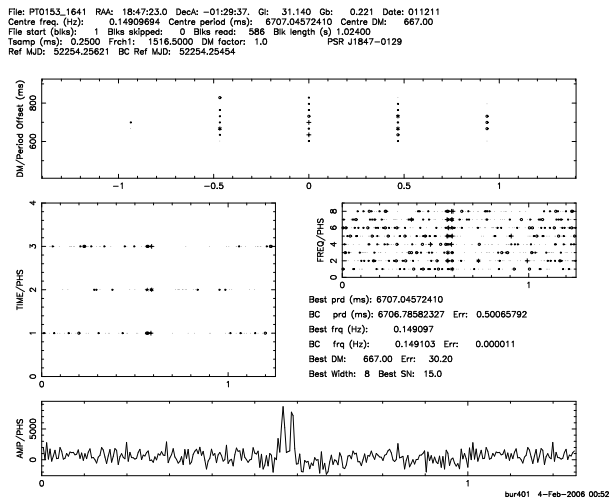}
\caption{Output plot of the {\ttfamily{pdm}} programme for the
high magnetic field pulsar J1847--0130. From the top, the panels
show a grey-scale of the S/N as a function of the trial values of P
and DM (top), the intensity of the pulsed signal as a function of the
pulse phase in three time sub-integrations (left) and in eight
frequency bands (right) and the integrated pulse profile (bottom)
obtained with the P and DM values maximising the S/N.}
\label{fig2}
\end{figure}
\end{center}

A typical output plot resulting from the {\ttfamily{pdm}} code is
shown in Figure \ref{fig2} where the signal of the high-magnetic
field pulsar PSR J1847-0130 has been searched around its nominal
period and DM in the ranges indicated on the axes of the top
grey-scale showing the S/N (the darkest the highest) as a function
of the trial values of P and DM. The central panel on the left
shows the intensity of the pulsed signal as a function of the
pulse phase in three sub-integrations three minutes long each. The
central panel on the right shows the folded time series in eight
frequency sub-bands. The bottom panel is the integrated pulse
profile obtained summing the sub-integrations and sub-bands using
the apparent (topocentric) P and DM which give the maximum S/N.
These values, as well as the value of the period as seen in the
solar system barycenter (BC) and the best S/N are reported on the
plot.

For each of our targets we have split the total range of the searched
parameters in $\sim 40$ sub-ranges, producing and visually inspecting
a plot for each of them. A {\ttfamily{pdm}} plot has been considered
interesting -- i.e. representing a promising radio pulsar candidate --
if it showed a high enough S/N (above a threshold of 6), a well
defined peak in the bottom panel and clear linear trends in the
grey-scales of sub-bands and sub-integrations.

With the aim to search for signals from the AXPs with periods
different from the ones detected in X-rays and in the hope to discover
new radio pulsars, the data for all 13 beams have been analysed
searching for pulsations over a wide range of periods. To do so the
{\ttfamily{pmsearch}} code (see e.g. \citealt{mlc+01}) has been
used. This code is based on the analysis of the Fourier spectra of the
de-dispersed time series, obtained, for each different DM, by summing
with the appropriate delay the 96 frequency channels in which the
total band is splitted. The ``Tree'' algorithm \citep{tay74} used for
the de-dispersion works in such a way that the maximum DM explored in
this case is 10495 pc cm$^{-3}$, corresponding to a broadening of the
pulse in a single channel equal to the sampling time doubled five
times. 
{\ttfamily{Pmsearch}} includes also a time domain search algorithm
(Fast Folding Algorithm; \citealt{fsk+04}) sensitive to periods
greater than $\sim 3$ s, where the frequency resolution of the Fourier
spectra is very poor and the FFT is consequently less efficient.

{\ttfamily{Pmsearch}} is one of the blind search codes used for
the major pulsar surveys performed using the Parkes radio telescope
and led to the discovery of more than 700 new pulsars (see
e.g. \citealt{fsk+04} and references therein).

Finally a code sensitive to strong single dispersed pulses
\citep{cm03} has been applied to the 13 beams of each pointing with
the purpose to discover signals similar to those seen in the class of
the Rotating RAdio Transients (RRATs, \citealt{mll+06}). The aim was
to test if AXPs have RRAT-like emission and to search single pulses
from nearby sources.

\section{Results}

\subsection{Upper limits on the radio pulsation from the observed sample of AXPs}
\label{sec:results}

No radio pulsation has been found in the four AXPs observed. An upper
limit for pulsed radio flux can be estimated by using equation \ref{eq1}
(e.g. \citealt{mld+96}), representing the minimum detectable flux
density from a pulsar of period P.
\begin{equation}
S_{min}= \epsilon n_{\sigma}
\frac{T_{ sys}+T_{sky}}{G\sqrt{N_p\Delta t \Delta \nu_{{MHz}}}}
\sqrt{\frac{W_e}{P-W_e}} \qquad {\rm mJy}
\label{eq1}
\end{equation}
Here $n_{\sigma}$ is the minimum S/N considered (in this case 6.0),
$T_{sys}$ and $T_{sky}$ the system noise temperature and the sky
temperature in K respectively, $G$ the gain of the radio telescope
in K/Jy, $\Delta t$ is the integration time in seconds, $N_p$ the
number of polarisations and $\Delta \nu_{MHz}$ the total bandwidth
in MHz. $\epsilon$ is a factor $\sim 1.4$ accounting for
sensitivity reduction due to digitisation and other losses. $W_e$
is the effective width of the pulse:
\begin{equation}
W_e=\sqrt{W^2+\delta t^2+\delta t_{DM}^2 + \delta t_{scatt}^2}
\label{eq2}
\end{equation}
Its  value depends on the intrinsic pulse width $W$, on the sampling
time $\delta t$ and on the broadening of the pulse introduced both by
the dispersion of the signal in each 3 MHz channel ($\delta t_{DM}$)
and by the scattering induced by inhomogeneities in the ISM
($\delta t_{scatt}$).

Prior to digitisation, the signals are high-pass filtered by
hardware in order to eliminate slowly varying baseline levels.
This introduces, depending on the period of the pulsar, a further
degrading factor of the sensitivity of the data acquisition
system. Following \citet{mlc+01} we estimate a degradation of
$\sim 40\%$ for periods around $6-7$ s and of $\sim 50\%$ for
periods around $10-12$ s.

Adopting the appropriate numbers for the central beam of the multibeam
receiver ($G=0.735$ Jy/K, $T_{sys}=23$ K, $N_p=2$), an intrinsic pulse
width of 5\% of the pulse period and the values reported in Table
\ref{tab2} we obtain, depending on the source and considering the
aforementioned degrading factors, upper limits in the range $\sim
0.06 - 0.1$ mJy.

\begin{table*}
\begin{center}
\begin{tabular}{lcc|rccrc}
\hline
Name & P range & DM$_{max}$ & DM$_{max}^{gal}$ & $\delta t_{scatt}^{gal}$  & $\delta t$ & $T_{sky}$ & $S_{min}$ \\
 & (s)& (pc cm$^{-3}$) & (pc cm$^{-3}$) & (ms) & (ms) & (K) & mJy \\
\hline
\ee    & \ 6.0438 --  \ 7.2524 &\ 62754 &  682.4 &   0.6 & 1.2 &  5.4 & 0.11 \\
\rxj   & \ 9.6247 -- 12.3734 &   107069 & 1695.5 & 127.6 & 1.0 & 13.1 & 0.06 \\
\kes   &  10.3026 -- 13.2451 &   114611 & 1581.5 & 118.9 & 1.0 & 11.8 & 0.06 \\
\axj   & \ 6.0999 --  \ 7.8426 &\ 67861 & 1494.4 &  88.8 & 1.0 & 12.9 & 0.06 \\
\hline
\end{tabular}
\caption{For each observed AXP we report here the period range and
  maximum DM explored in the {\ttfamily{pdm}}search (columns 2 and 3)
  and the values used to calculate (according to equation \ref{eq1})
  the upper limit on the flux densities reported in column 8: column 4
  reports the maximum DM compatible with the pulsar being in the
  Galaxy (chosen as the maximum between the values given by the
  \citet{tc93} and the \citet{cl02} models), column 5 the expected
  broadening induced by the scattering in the ISM at 1.4 GHz, column 6
  the sampling time and column 7 the sky temperature at 1.4 GHz in
  the direction of the source.}
\label{tab2}
\end{center}
\end{table*}

For 1E1048.1-5937 the flux density limit would have been $\sim 0.04$
mJy but the observations have been mistakenly performed with an offset
with respect to the true position of the source of 8 arcminutes in
right ascension. Assuming a Gaussian beam shape of width 14.4
arcminutes \citep{hfs+04}, a further degrading factor $\sim 2.5$
caused by the non-uniform beam response must hence be added for estimating
the flux upper limit for this object.

Folding the time series for \ee with a trial period of 6.16 s and
a DM of 334 pc cm$^{-3}$, the pulsar J1058-5957 \citep{kbm+03} has
been detected through its tenth sub-harmonic, confirming the
reliability of the detection algorithm used.

\subsection{Discovery of two new radio pulsars}
\label{sec:pulsar}

\begin{figure}
\begin{center}
\includegraphics[width=8truecm]{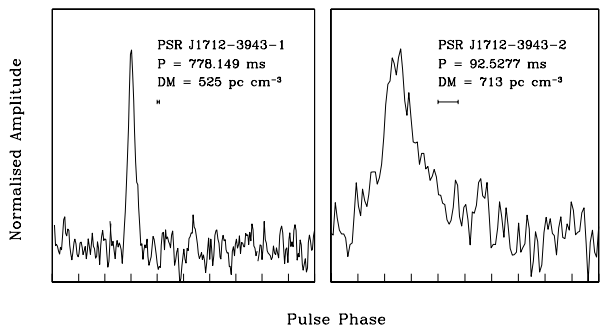}
\caption{Mean 1374-MHz pulse profiles of the newly discovered pulsars
\newpsr-1 (left) and \newpsr-2 (right) obtained by adding the data of
the discovery and confirmation observations. The maximum of each
profile is placed at phase 0.3 (the range on the x-axis of each panel
being 0 to 1). For each profile, the pulsar name, period and DM are
given. The small horizontal bar drawn under the DM indicates the
effective resolution of the profile, calculated by adding the bin
size, the sampling time and the effects of interstellar dispersion in
quadrature.}
\label{fig3}
\end{center}
\end{figure}

Analysis of the data collected in all the 13 beams of the Parkes
multibeam receiver used during this work led to the serendipitous
discovery of two new radio pulsars, both detected in the same
beam. Since their precise celestial coordinates are yet to be
determined, their provisional names are \newpsr-1 and \newpsr-2.  The first
pulsar has a spin period of 0.78 s and a dispersion measure of 525 pc
cm$^{-3}$, while the second source has a period of 92.5 ms and a DM of
713 pc cm$^{-3}$ (see Table \ref{tab3}). Given the large difference in
the DMs, we can exclude a physical association between the two
pulsars. An immediate confirmation of the detection of the new pulsars
has been possible thanks to the duplication of the pointings performed
on three of our four targets. 
The average pulse profiles of the new pulsars, obtained summing the
data relative to the detection and confirmation observations, are
shown in Figure \ref{fig3}.

We note that at a distance of 12.2 arcmin from the centre of the beam
where the new pulsars have been found there is the central compact
object (CCO) 1WGA J1713.4--3949 in the supernova remnant G347.3--0.5
\citep{lsg+03}. This source could be associated with the 92.5 ms
pulsar \newpsr-2: the distance of the CCO from the nominal detection
position of the radio pulsar is, in fact, compatible with the width of
the beam of the telescope and the DM derived distance of the pulsar D
= 7.7 kpc (using the \citealt{cl02} model, for which a typical 20\%
uncertainty is expected) is in agreement with the distance of the
supernova remnant ($6 \pm 1$ kpc; \citealt{sgd+99}). If this pulsar is
indeed the same object as 1WGA J1713.4--3949, we can give a rough
estimate of the spin period first derivative $\dot{P}$ that results
$\sim 10^{-14}$ by imposing that the observed X-ray luminosity $L_x =
6\times 10^{34}$\,erg\,s$^{-1}$ \citep{lsg+03} is, as for the bulk of isolated
millisecond pulsars detected in X-ray and radio, 0.1\% of the spin
down energy $\dot{E}$ \citep{bt97,pccm02}.
Unfortunately the period first derivative, calculated using our two
detections of \newpsr-2 and assuming that its true position is that of
the CCO, is totally unconstrained ($\dot{P}_{obs}=[1.9 \pm 2.7] \times
10^{-9}$) and cannot be compared with the value estimated above. The
observations where \newpsr-2 has been detected are infact only few hours
apart and the periods measured in the two datasets are identical, at
1-$\sigma$ level.
Using the $\dot{P}$ derived from the X-ray luminosity we can also
calculate the putative age of the pulsar $\tau = P/2\dot{P}\sim 100$
kyr, only a factor of $\sim 2$ larger than the age of G347.3--0.5
($\ltsim 40$ kyr; \citealt{sgd+99}), a good agreement given the
crudeness of the estimates. On the other hand, considerations on the
flux density of \newpsr-2 weaken the probability of an association
with 1WGA J1713.4--3949: using equation \ref{eq1} and considering the
12.2 arcmin offset, infact, we obtain that \newpsr-2 would have a flux
at 1.4 GHz $S_{1400} \sim 2$ mJy at the CCO position. Targetted radio
observations of the supernova remnant hosting 1WGA J1713.4--3949, on
the contrary, gave negative results providing a very tight upper limit
of 0.06 mJy \citep{cpkm02}. Since radio frequency interferences may
have hampered the detection of the 92.5 ms pulsar in Crawford's et
al. (2002) single pointing, further observation are nevertheless
necessary in order to establish if an association with the CCO in
G347.3--0.5 (or other close-by high energy sources) exists and to
determine whether the millisecond pulsar is young or recycled. In this
second case, given the relatively long spin period, \newpsr-2 could be
a mildly recycled pulsar.

The {\ttfamily{pmsearch}} code also detected 18 previously known
pulsars present in the pointed areas. A list of the pulsars found
during this search (new and known), with detection period, DM, beam
position and S/N (calculated in the time domain) is given in Table
\ref{tab3}.

\begin{table*}
\begin{center}
\begin{tabular}{lccccr}
\hline
Pulsar & P & DM & RAJ & DECJ & S/N \\
 & ms & \scriptsize{pc/cm$^3$} & h m s & $^\circ\,^\prime\,^{\prime\prime}$ &  \\
\hline
J1712--3943-1 & 778.149(1)  & 525(3) & 17:12:35 & -39:43:14 & 16.6 \\
J1712--3943-2 &  92.5277(1) & 713(3) & 17:12:35 & -39:43:14 & 12.1 \\
\hline
J1054--5943 & 346.9090(4) & 332(2) & 10:54:16 & -59:53:03 & 25.8 \\
J1055--6022 & 947.551(3)  & 693(7) & 10:56:11 & -60:18:24 & 12.8 \\
J1058--5957 & 616.271(1)  & 334(4) & 10:58:08 & -59:53:16 & 104.4 \\
J1103--6025 & 396.5869(5) & 278(2) & 11:04:03 & -60:18:09 & 26.2 \\
J1705--3936 & 854.482(2)  & 598(5) & 17:04:58 & -39:43:14 & 13.4 \\
J1705--3950 & 318.9408(3) & 202(2) & 17:04:58 & -39:43:14 & 47.6 \\
J1707--4053 & 581.017(2)  & 348(5) & 17:07:30 & -40:34:00 & 10.3 \\
J1713--3949 & 392.4514(5) & 342(2) & 17:12:36 & -39:43:14 & 14.4 \\
J1839--0459 & 585.319(1)  & 235(3) & 18:39:22 & -04:56:11 & 27.8 \\
J1841--0425 & 186.1478(1) & 325(2) & 18:40:21 & -04:31:00 & 29.4 \\
J1841--0524 & 455.7312(7) & 283(3) & 18:42:17 & -05:21:23 & 12.8 \\
J1842--0359 & 1839.95(1)  & 187(9) & 18:41:19 & -04:05:26 & 9.7 \\
J1843--0459 & 754.963(2)  & 434(4) & 18:43:16 & -04:56:11 & 142.6 \\
J1844--0244 & 507.725(1)  & 421(3) & 18:44:53 & -02:56:42 & 17.1 \\
J1844--0256 & 272.963(1)  & 822(4) & 18:44:53 & -02:56:42 & 21.0 \\
J1844--0433 & 911.028(3)  & 128(6) & 18:44:15 & -04:30:47 & 160.0 \\
J1844--0538 & 255.7013(4) & 413(2) & 18:44:16 & -05:21:32 &  9.7 \\
J1845--0316 & 207.6358(1) & 503(2) & 18:45:51 & -03:21:53 & 31.1 \\
\hline
\end{tabular}
\caption{New (first two rows) and redetected pulsars. From left to
  right pulsar name (for the new pulsars is only a provisional
  name given the position uncertainty), detection period and
  dispersion measure, right ascension and declination (in coordinates
  J2000) of the centre of the beam containing the pulsar and detection
  signal-to-noise ratio. The number in parentheses are the 2-sigma
  errors on the last quoted digit.}
\label{tab3}
\end{center}
\end{table*}

\begin{figure}
\begin{center}
\includegraphics[width=8truecm]{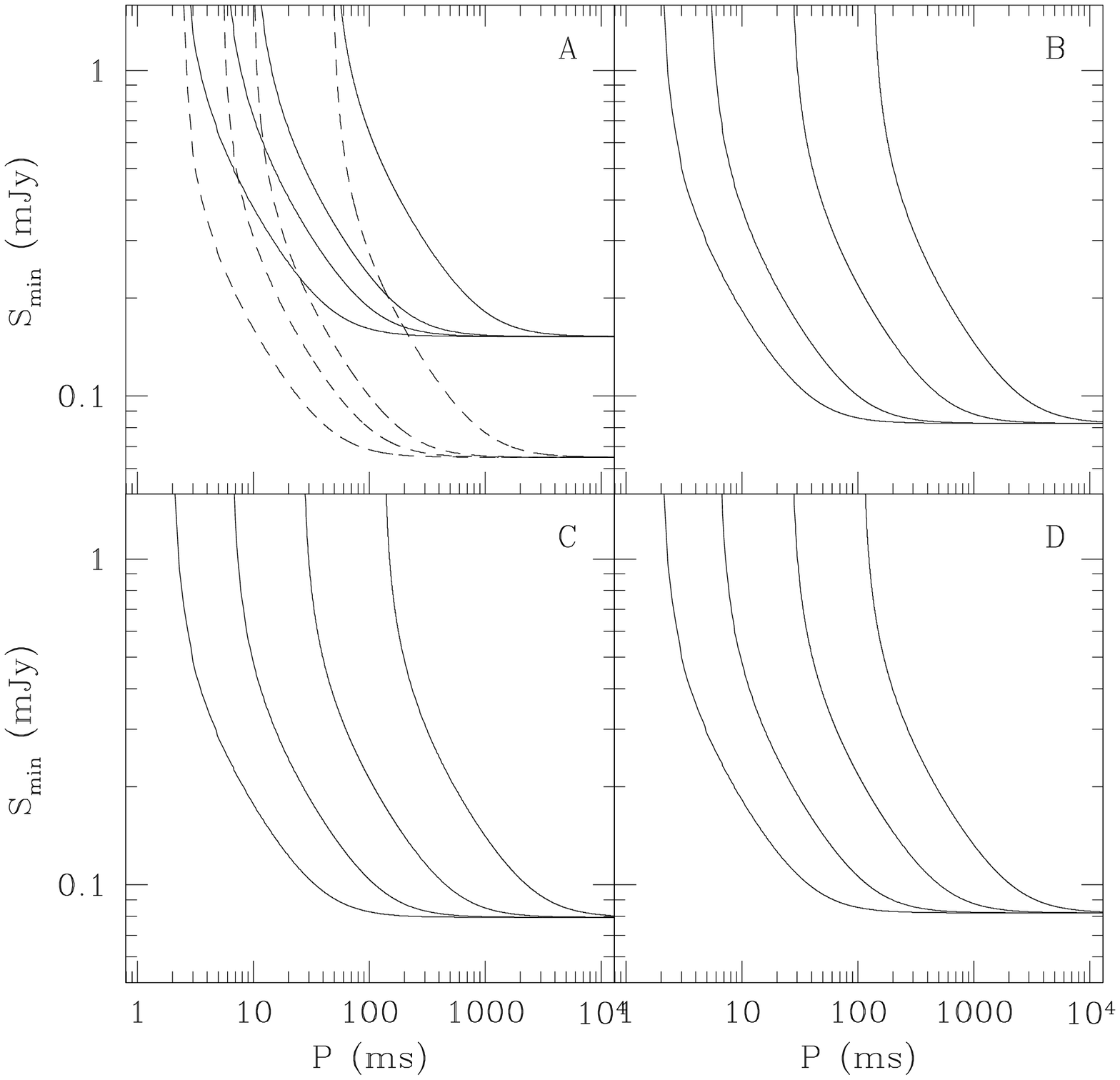}
\caption{Flux density limits for the four observed AXPs, and
  surrounding beams, as a function of the spin period and for
  different values of the DM (from bottom to top: 0, 500, 1000 and
  5000). Panels A, B, C and D refer respectively to the pointings
  centered on \ee, \axj, \kes and \rxj. For panel A the solid lines
  refer to the flux limit for the pulsation from the AXP (taking
  into account the 8' offset of the central beam with respect to the
  source position) while the dashed line is the flux limit for the
  blind search in all the 13 beams.}
\label{fig4}
\end{center}
\end{figure}

None of the pulsed signals found during this blind search could
arise from any of the AXPs that we observed, because of the
discrepancy in position. Figure \ref{fig4} shows, for four
different values of the DM, the upper limits on the flux density
at 1.4 GHz reached for the four pointed regions of the sky as a
function of the spin period. These limits have been estimated
using equation \ref{eq1} with $n_{\sigma}=8$ and multiplying the
limit obtained by a factor $\sim 2$ taking into account losses due
to the search technique (see \citealt{mlc+01}).
Note that the $n_{\sigma}$ threshold adopted here is higher with
respect to the one used in the previous step of the analysis: this is
because, given the huge amount of radio frequency interferences (RFI)
with which we have to deal when doing a blind periodicity search
(especially at low S/N), we cannot safely state as a good candidate a
suspect with S/N less than, usually, 8. The RFI problem is
not so important when folding the time series at a specific value of
the period, hence in that case we can be less conservative using a
lower threshold for the candidate selection and, consequently, for the
$n_{\sigma}$ to adopt in the $S_{min}$ calculation.

\begin{figure}
\begin{center}
\includegraphics[width=8truecm]{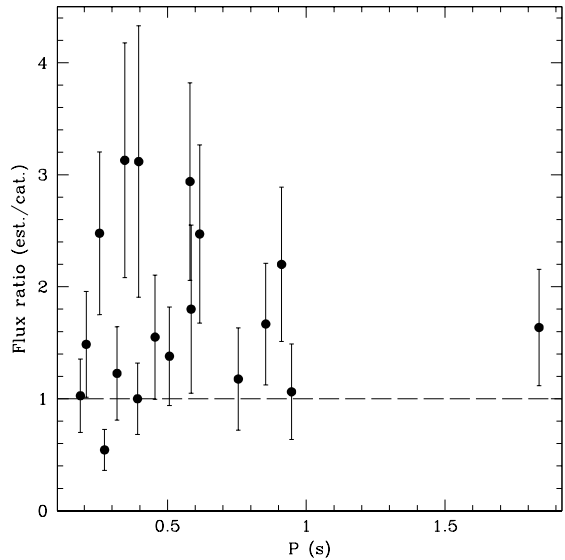}
\caption{Ratio between the estimated (through equation \ref{eq1}) and
  the catalogued value of the flux density of the 18 redetected
  pulsars. The errorbars are drawn assuming, besides the errors on the
  measured values, a 20\% error on the estimated fluxes. The mean
  value of the ratio calculated over the 18 sources is $1.15 \pm 0.3$
  (2-$\sigma$ error).}
\label{fig5}
\end{center}
\end{figure}

In order to verify the reliability of the quoted flux density limits,
we can compare the catalogued fluxes of the 18 redetected pulsars
(obtained from the ATNF Pulsar Catalogue:
{\ttfamily{http://www.atnf.csiro.au/research/pulsar/psrcat/}}) with
those estimated putting in equation \ref{eq1} their detection periods,
DMs and S/Ns (or mean S/N, for pulsars with multiple detections) and
considering the offset of the detection beam with respect to the true
position of the pulsars. The ratio between the estimated and the
catalogued values of the flux densities are shown, as a function of
the rotational period, in figure \ref{fig5}. Note that the S/N used to
estimate the flux density, may sensibly vary from one observation to
another (e.g. because of the different response of different beams in
which the pulsar is detected, or because of RFI), leading to a large
uncertainty in the estimate of the flux density. Averaging the S/Ns
from different observations would lead to a better match between the
estimated and catalogued fluxes.

Finally no new RRAT-like signal has been found in the single pulse
search over the 52 observed regions, while two previously known
pulsars (J1705--3950 and J1844--0433) have been detected also through
their single pulses.

\section{Discussion and Conclusions}
\label{sec:conclu}

No radio pulsation has been found from the four southern AXPs
observed, down to a limit of $S_{min} \sim 0.1$ mJy (Table
\ref{tab2}).

\begin{figure}
\begin{center}
\includegraphics[width=8truecm]{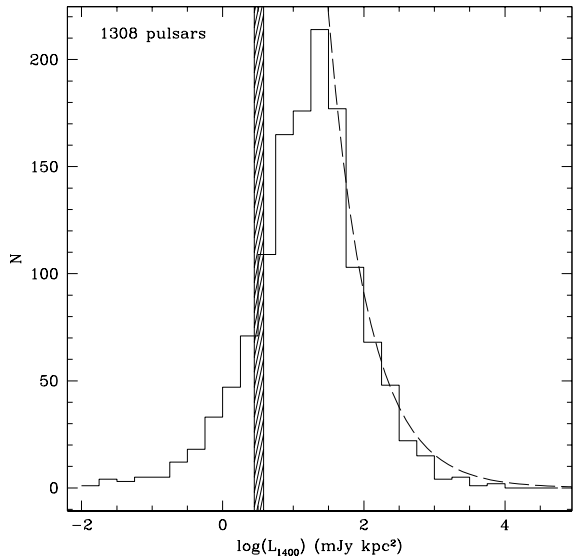}
\caption{Distribution in luminosity at 1400 MHz of the observed pulsar
  population of the galactic field (solid line) and of the intrinsic
  population (dashed line; \citealt{lfl+06}). The shaded region
  indicates the luminosity limits reached in the present search.}
\label{fig6}
\end{center}
\end{figure}

Comparing the upper limits on the luminosity at 1400 MHz (defined as
$L_{1400} = S_{min} \times D^2$) for our targets with the luminosity
of the observed radio pulsar population in the Galactic field (Figure
\ref{fig6}), we note that the limits reached by our search are a
factor of six lower than the median of the population. The values
obtained for our sample are in fact in the range 2.8 -- 3.8 mJy
kpc$^2$ (shaded region in Fig. \ref{fig6}) while the median of the
distribution in luminosity of the observed radio pulsars in our Galaxy
is 18.0 mJy kpc$^2$. If the luminosity function of AXPs reflects that
of observed canonical pulsars we can calculate that the probability
for all our four targets to have luminosities below our limits is only
$5\times 10^{-4}$.

Similar results are obtained comparing the upper limits on
$L_{1400}$ for our sample with the luminosity distribution of only
young pulsars (with characteristic ages $< 10^4$ yr) or long
period pulsars (with P $>$ 3 s).

A comparison with the few high magnetic field radio pulsars
(having magnetic field strengths above the quantum critical field
line; Figure \ref{fig1}) is not statistically significant, having
only four objects in this class. However we note that all their
luminosities, as well as the radio luminosity of \xte
\citep{crh+06}, are greater than the limits obtained for the AXPs.

Note however that, if we compare our results with the {\it{intrinsic}}
luminosity distribution of the radio pulsars (dashed line in figure
\ref{fig6}; \citealt{lfl+06}), we obtain a probability of $\sim 76$\%
that our observations are not deep enough to detect a radio pulsed
signal from our targets.

In considering the causes of the non detection of a radio pulsed
signal from our four targets, besides the luminosity bias, we must
take into account the possibility that, although the X-ray beam is
pointing toward us, the radio beam, usually narrower, is not. Assuming
a pulse duty cycle of $\sim 5\%$ (hence a beam semi-aperture $\geq
9^{\circ}$), typical of long period pulsars and similar to that of the
radio pulsed signal detected by \citet{crh+06} in the transient AXP
\xte ($\sim 4\%$ at 1.4 GHz), we can calculate, following
e.g. \citet{bbp+03}, that the probability that such a narrow radio
beam misses the earth is $\leq 77\%$. The composite probability that
the beams of all four AXPs are not pointing toward us is hence $\leq
34\%$.

The non detection of RRAT-like bursts from any of these AXPs, despite
our long exposures, seems to weaken the hypothesis that RRAT bursts
might be related to the short bursts observed from the magnetars
leaving us with other plausible conjectures of a relation with other
classes of neutron stars such as middle aged radio pulsars displaying
giant pulses \citep{rbg+06,wsrw06} or with X-ray Dim Isolated Neutron
Stars \citep{ptp06}.

The only case of a detection of radio pulsations from an AXP concerns
the only confirmed transient magnetar \xte\, \citep{crh+06}.  Radio
emission from this source is strongly related with the occurrence of
an outburst of its X-ray emission \citep{hgb+05}, as well as an IR
enhancement \citep{rti+04}. Furthermore, whereas the X-ray flux is
decaying exponentially with timescale of a few hundreds days
\citep{gh05}, \xte\, radio emission is still on more than 3 years
after the X-ray outburst. Interestingly the sole other possible
transient AXP is the candidate \axj, one of our targets. Our radio
observations of this source were performed more than six years after
its possible X-ray outburst occurred in 1993, hence unfortunately
nothing can be safely concluded from our upper limits, in favor or
against the possible radio and X-ray correlation during the outbursts
of this source. However, assuming that \axj experienced, after the
X-ray outburst, a phase of radio emission similar to that of \xte, our
null detection implies that the fading of the radio emission has a
time scale of the order of few years: in particular, if \axj at the
onset of its putative radio emission phase had a similar luminosity as
\xte, this would imply a decrease in $L_{1400}$ of a factor of $\sim
20$ over six years.

\section*{Acknowledgments}
MB, AP, LB, TDS and NDA received support from the Italian Ministry of
University and Research (MIUR) under the national program {\it
PRIN2005 2005024090\_001}. NR is supported by an NWO post-doctoral fellow. The
Parkes radio telescope is part of the Australia Telescope which is
funded by the Commonwealth of Australia for operation as a National
Facility managed by CSIRO.



\bsp
\label{lastpage}

\end{document}